\documentclass[aps,preprint,nofootinbib,showpacs,superscriptaddress,groupedaddress]{revtex4}
\usepackage{graphicx}
\usepackage{epstopdf}

\usepackage{amsmath,amsthm,latexsym,amssymb,amsfonts,epsfig}
\usepackage{fontenc,dsfont,layout}
\usepackage{hyperref}
\usepackage{psfrag}
\usepackage[cp1250]{inputenc}

\usepackage[usenames,dvipsnames]{xcolor}

\numberwithin{equation}{section}

\begin{document}
\title{\boldmath Saddle point inflation from higher order corrections to Higgs/Starobinsky inflation}

\author{Micha{\l} Artymowski}
\affiliation{Institute of Physics, Jagiellonian University\\
{\L}ojasiewicza 11, 30-348 Krak{\'o}w, Poland}
\author{Zygmunt Lalak}
\affiliation{Institute of Theoretical Physics, Faculty of Physics, University of Warsaw\\
 ul. Pasteura 5, 02-093 Warsaw, Poland}
\author{Marek Lewicki}
\affiliation{Institute of Theoretical Physics, Faculty of Physics, University of Warsaw\\
 ul. Pasteura 5, 02-093 Warsaw, Poland}



\begin{abstract}
We explore two saddle point inflationary scenarios  in the context of higher order corrections related to different generalisations of general relativity. Firstly, we deal with Jordan frame Starobinsky potential, for which we identify a portion of a parameter space of inflection point inflation, which can accommodate all the experimental results. Secondly, we analyse Higgs inflation and more specifically the influence of non-renormalisible terms on the standard quartic potential. All results were verified with the PLANCK 2015 data.
\end{abstract}


\date{\today}

\maketitle

\tableofcontents

\section{Introduction}

Cosmic inflation \cite{Lyth:1998xn,Liddle:2000dt,Mazumdar:2010sa} is a theory of the early universe which predicts cosmic acceleration and generation of seeds of the large scale structure of the present universe. It solves problems of classical cosmology and it is consistent with current experimental data \cite{Ade:2013uln}. It is instructive to study minimal models of inflation which have the form of small modifications of the canonical models describing features of the visible Universe. Two most prominent examples of such models are, at the moment,  $f(R)$ theories of gravity models of Higgs inflation.

The first theory of inflation was the Starobinsky model \cite{Starobinsky:1980te,Barrow:1988xh}, which is an $f(R)$ theory \cite{DeFelice:2010aj} with $R + R^2/6M^2$ Lagrangian density. In such a model the acceleration of space-time is generated by the gravitational interaction itself, without a need to introduce any new particles or fields. The embedding of Starobinsky inflation in no-scale SUGRA has been discussed in Ref. \cite{Ellis:2015xna}. The Starobinsky inflation can be described as a special case of the Brans-Dicke theory \cite{Brans:1961sx} and its predictions are consistent with the so-called Higgs inflation \cite{Bezrukov:2007ep}, in which inflation is generated by the scalar field non-minimally coupled to gravity. Recently the whole class of generalisations of the Starobinsky and Higgs inflation have been discussed in the literature \cite{Codello:2014sua,Ben-Dayan:2014isa,Artymowski:2014gea,Artymowski:2014nva,Sebastiani:2013eqa,Motohashi:2014tra,Salvio:2015kka,Hamada:2014wna}, also in the context of the higher order terms in Starobinsky Jordan frame potential \cite{Broy:2014sia,Kamada:2014gma,Artymowski:2015mva}.
\\*

On the other hand, Higgs inflation allows one to generate inflation close to the Higgs sector. The price is the non-minimal coupling to gravity and often additional interactions which allow one to reproduce all features of realistic inflationary scenarios. In particular, corrections to the model can often be represented by higher order scalar operators. This reminds one of the situation known from $f(R)$ theories, where higher order corrections to the f(R) function seem to be unavoidable. {Usually one assumes that there is a part of the Starobinsky/Higgs inflationary potential where higher order terms are subdominant, which creates an inflationary plateau long enough to support successful inflation. Nevertheless one can find a part of parameter space where higher order corrections give significant contribution to the potential for relatively small values of field, which makes the plateau region too short to generate cosmic inflation with at least 60 e-foldings. In this case a saddle point inflation generated by the higher order terms may be the only chance to obtain a successful inflationary model, which is the issue we investigate in this paper.}
\\*

In \cite{Artymowski:2015mva} we proved that the Starobinsky potential with higher order corrections (i.e. higher powers of the Jordan frame Starobisnky potential) can have a saddle point for some $\phi$ on the plateau. This requires a certain relation between parameters of the model and leads to the existence of the Starobinsky plateau, a step slope of exponential potential and a saddle point in between. In this paper we perform the detailed analysis of such a model in the context of low-scale inflation and generation of primordial inhomogeneities, especially in the context of very big values of higher order terms. We apply the same approach to the Higgs inflation. In this case the motivation is even more natural: higher order corrections to the scalar potential (such as $\psi^6$ or $\psi^8$, where $\psi$ is a scalar field, but not necessarily the Higgs field itself) are considered to be suppressed by the Planck scale. Consequently for sufficiently high values of field $\psi$ one has to take into account the influence of higher order terms. In particular higher order terms coefficients could be fine-tuned to create a saddle point (or deflation point) in the Einstein frame scalar potential. {In Ref \cite{Artymowski:2015pna} we have also investigated the saddle-point inflation in $f(R)$ theory generated by higher order corrections to the Starobinsky model. Note that in this paper we investigate the issue of higher order corrections to Jordan frame scalar potentials, not to the $f(R)$ function itself.}
\\*

In what follows we use the convention $8\pi G = M_{p}^{-2} = 1$, where $M_{p}= 2.435\times 10^{18}GeV$ is the reduced Planck mass.
\\*

The outline of the paper is as follows. In Sec. \ref{sec:Infl} we introduce the form of the potential in modified Strobinsky case and we analyse its features. In Sec. \ref{sec:infl2} we analyse the saddle point inflation in this model and the evolution of primordial inhomogeneities.
In Sec \ref{sec:higgs} we investigate the saddle point within the Higgs inflation scenario.
Finally, we conclude in Sec. \ref{sec:concl}. 


\section{Starobinsky-like potential with a saddle point} \label{sec:Infl}

\subsection{Jordan frame analysis}

Let us consider a Brans-Dicke theory in the flat FRW space-time with the metric tensor of the form $ds^2 = -dt^2 + a(t)^2(d\vec{x})^2$. {Any $f(R)$ theory can be expressed in terms of the auxiliary field $\varphi:=F(R):=f'(R)$ with the Jordan frame scalar potential $U=\frac{1}{2}(R F-f)$.} The Jordan frame action is of the form
\begin{equation}
S =  \int d^4x \sqrt{|g|}\left(\varphi R - U(\varphi)\right) + S_{\text{m}}\, , \label{eq:actionJ}
\end{equation}
where $S_{\text{m}}$ is the action of matter fields. {In the context of inflation one can assume that the energy density of the universe is fully dominated by the inflaton, which gives $S_m = 0$.} Then, for the homogeneous field $\varphi$ the field's equation of motion and the first Friedmann equation become \cite{DeFelice:2010aj}
\begin{eqnarray}
\ddot{\varphi} + 3H\dot{\varphi} + \frac{2}{3}(\varphi U_\varphi - 2U) &=& 0\ ,\label{eq:motionBD}\\
3\left(H + \frac{\dot{\varphi}}{2\varphi}\right)^2 &=& \frac{3}{4}\left(\frac{\dot{\varphi}}{\varphi}\right)^2 + \frac{U}{\varphi}\, , \label{eq:FriedBD}
\end{eqnarray}
where $U_\varphi:=\frac{dU}{d\varphi}$. Let us note that for $\varphi = 1$ one recovers general relativity (GR). Thus the $\varphi=1$ will be denoted as the GR vacuum.
\\*

The Starobinsky inflation is a theory of cosmic inflation based on the $f(R) = R + R^2/6M^2$ action, which can be generalized into Brans-Dicke theory with general value of $\omega_\text{\tiny BD}$. The Jordan frame potential of the Starobinsky model takes the following form
\begin{equation}
U_{S} = \frac{3}{4}M^2\left(\varphi-1\right)^2 \, , \label{eq:StarobinskyP}
\end{equation}
where $M$ is a mass parameter, {and its} value comes from the normalisation of the primordial inhomogeneities. For $\omega = 0$ one finds $M\simeq 1.5\times 10^{-5}$. The Starobinsky potential is presented in Fig. \ref{fig:Saddle}. In this paper we consider the extension of this model motivated by Ref. \cite{Broy:2014sia} and partially analysed in Ref. \cite{Artymowski:2015mva}, namely
\begin{equation}
U = U_S\left(1 - \lambda_1 U_S + \lambda_2 U_S^2\right) \, , \label{eq:StarobinskyGeneral}
\end{equation}
where $\lambda_1$, $\lambda_2$ are numerical coefficients. In order to avoid $U\to-\infty$ for $\varphi \to \infty$ we assume that $\lambda_2>0$. We want to keep $\lambda_i$ terms as higher order corrections {(i.e. we want to remain in perturbative regime of the theory)}, thus we require $\lambda_1 M^2\ll1$ and $\lambda_2 M^4 \ll \lambda_1 M^2$. As we will show the assumed range of parameters will satisfy these conditions.

\subsection{Einstein frame analysis} \label{sec:EinAn}

The gravitational part of the action may obtain its canonical (minimally coupled to $\varphi$) form after transformation to the Einstein frame. Let us assume that $\varphi>0$. Then for the Einstein frame metric tensor
\begin{equation}
\tilde{g}_{\mu\nu}=\varphi g_{\mu\nu}\, , \qquad d\tilde{t}=\sqrt{\varphi}dt\, ,\qquad\tilde{a} = \sqrt{\varphi}a
\end{equation}
one obtains the action of the form of 
\begin{equation}
S[\tilde{g}_{\mu\nu},\phi] = \int d^4x \sqrt{-\tilde{g}}\left[ \frac{1}{2}\tilde{R} - \frac{1}{2}\left( \tilde{\nabla}\phi \right)^2 - V(\phi)\right] \, ,
\end{equation}
where $\tilde{\nabla}$ is the derivative with respect to the Einstein frame coordinates, {$\phi = \sqrt{3/2}\log\varphi$, $V(\phi)=U/\varphi^2$ at $\varphi = \varphi(\phi)$} and $\tilde{R}$ is the Ricci scalar of $\tilde{g}_{\mu\nu}$. The GR vacuum appears for $\phi=0$. Let us define the Einstein frame Hubble parameter as
\begin{equation}
\mathcal{H}:=\frac{\tilde{a}'}{\tilde{a}}\, ,\qquad \text{where} \qquad \tilde{a}' := \frac{d\tilde{a}}{d\tilde{t}} \, . \label{eq:HEin}
\end{equation}
Then for $\rho_M=P_M=0$ the first Friedmann equation and the equation of motion of $\phi$ are following
\begin{equation}
3\mathcal{H}^2 = \frac{1}{2}\phi'^2 + V(\phi) \, , \qquad \phi'' + 3\mathcal{H}\phi' + V_\phi = 0 \, ,\label{eq:EOMEin}
\end{equation}
where $V_\phi = \frac{dV}{d\phi}$.
\\*

\subsection{Saddle point of the Einstein frame potential}

In Ref. \cite{Artymowski:2015mva} {we showed} that the Einstein frame potential of the model from Eq. (\ref{eq:StarobinskyGeneral}) {has} a saddle point at $\phi=\phi_s$ for $\lambda_1=\lambda_s$, where
\begin{equation}
\phi_s \simeq \frac{1}{\sqrt{6}}\log\left(\frac{10}{3 \lambda _1 M^2}\right) \, ,\quad \lambda_1 = \lambda_s \simeq \frac{5 \lambda _2^{3/5} M^{2/5}}{2^{1/5} 3^{2/5}} \, , \quad V_s \simeq \frac{3}{4} M^2 \left(1-\frac{15}{8}\left(24M^4\lambda _2\right)^{1/5}\right)\, , \label{eq:conditionlambda}
\end{equation}
where $V_s = V(\phi_s)$. Let us note that for $\phi \simeq \phi_s$ one obtains a saddle point inflation, which in principle could significantly decrease the scale of inflation. For $\lambda_1 < \lambda_s$ the $\lambda_1$ term is always subdominant  as compared to the other terms of $V(\phi)$ and it can be neglected in the analysis. Therefore, the potential has no stationary points besides the minimum at $\phi = 0$. On the other hand, for $\lambda_1 > \lambda_s$ one obtains a potential with a local maximum at the plateau, step slope for big $\phi$ and a minimum in between. This case was analysed in \cite{Artymowski:2015mva}. {The Eq. \ref{eq:conditionlambda} gives approximate value of $\lambda_s$ and $\phi_s$, but it predicts all derivatives of $V$ for $\phi=\phi_s$ and $\lambda_1 = \lambda_s$ to be of order of $M^{8/3}\lambda_1^{2/3}$. This is not consistent with the saddle-like point condition $V'(\phi_s),V''(\phi_s)\ll V'''(\phi_s)$ and therefore Eq. \ref{eq:conditionlambda}  is not accurate enough to calculate power spectra of primordial inhomogeneities.}
\\* 

The analysis presented in this subsection is done under several assumptions and approximations and therefore its accuracy is limited by them. For instance one finds $V_\phi(\phi_s) \sim \lambda_2^{2/5}M^{18/5}$ and $V_{\phi\phi}(\phi_s) \sim\lambda_2^{2/5}M^{8/5} $. We have assumed that $\lambda_1 \gg M^2\lambda_2$ and we have obtained $\lambda_1 \sim \lambda_2^{3/5}M^{2/5}$. Thus, the assumption is satisfied for $(\lambda_2 M^4)^{2/5} \ll 1$, which is also the condition for $V_{\phi\phi}(\phi_s)\ll 1$. One can see that the approximation used in this subsection is { far more accurate for the $V_{\phi} = 0$ condition than for the $V_{\phi\phi} = 0$ condition}. 

\begin{figure}[h]
\centering
\includegraphics[height=5.1cm,bb=0 0 288 210]{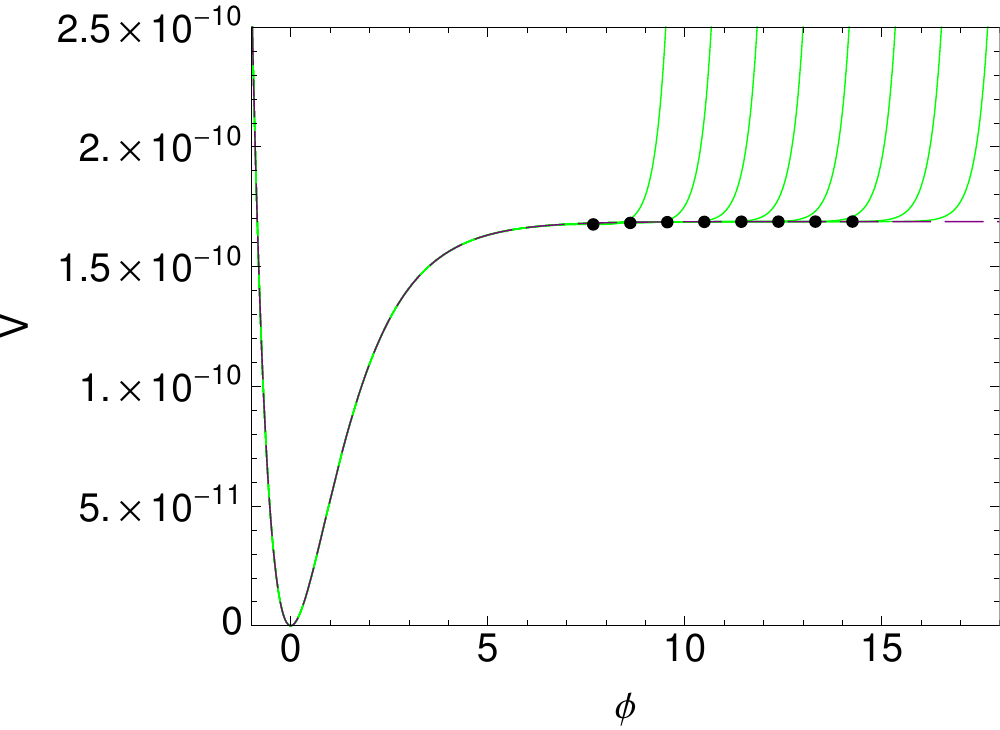}
\hspace{0.5cm}
\includegraphics[height=5.1cm,bb=0 0 260 196]{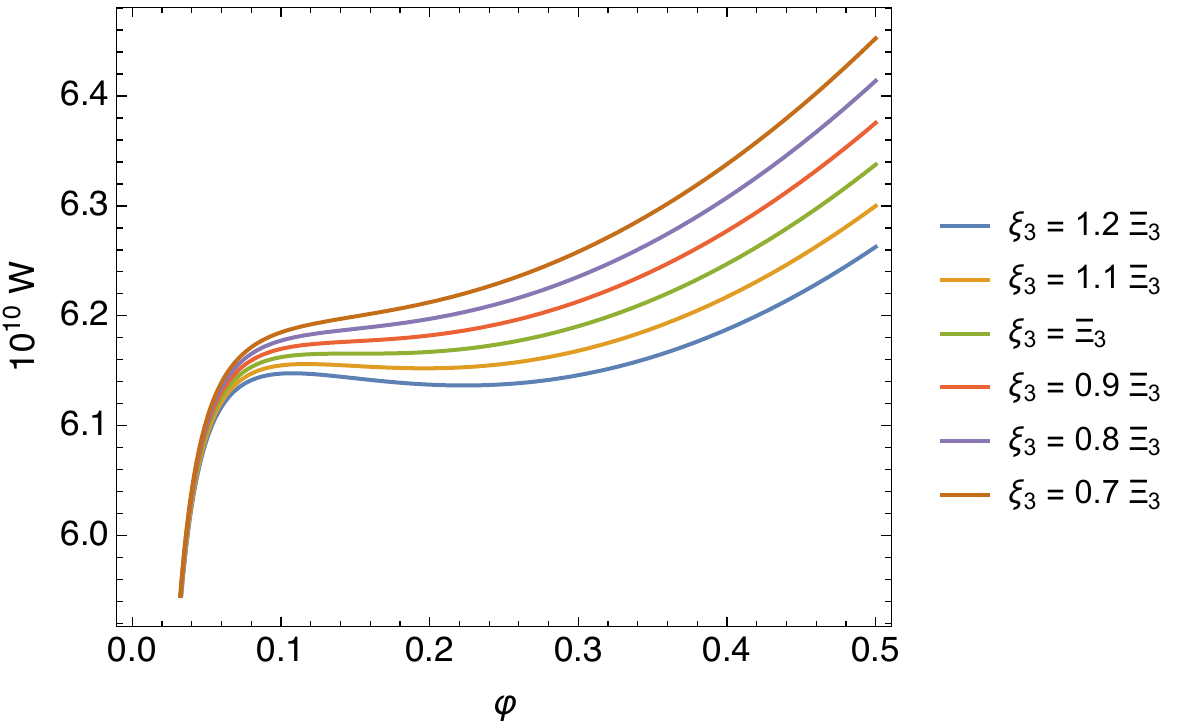}
\caption{\it  {Left panel: }$V(\phi)$ for Starobinsky inflation and for saddle point inflation (dashed purple line and green lines respectively). Black points represent $\phi = \phi_s$ for $\lambda_1 = 10^k$, where $k\in \{-5,-4,\ldots,2\}$ (smallest $\phi_s$ for biggest $\lambda_1$){. Right panel: $V(\Psi)$ with saddle point for $\lambda_8 = 1$, $\lambda_6 \propto -\lambda^{3/2}$, $\lambda\xi^{-2}\sim 2\times 10^{-9}$ and $\lambda \simeq \{0.15,0.4,1\}$ (dotted blue, dashed orange and solid red lines respectively). Dotted black line represents $V(\Psi)$ for $\lambda_6 = \lambda_8 = 0$. The saddle point inflation in this scenario will be analysed in Sec. \ref{sec:higgs}}}
\label{fig:Saddle}
\end{figure}


\section{Inflationary dynamics and primordial inhomogeneities} \label{sec:infl2}

\subsection{Inflation with and beyond the slow-roll approximation}

In this section we will discuss the slow-roll approximation and inflation in the Einstein frame. We want to investigate how non-zero values of $\lambda$ parameters deviate the inflation from the Starobinsky model. Let us assume that $\phi''\ll V_\phi$. Then one obtains
\begin{equation}
3\mathcal{H}\phi' + V_\phi \simeq 0\, , \qquad 3\mathcal{H}^2 \simeq V \, .\label{eq:SReoms}
\end{equation}
This approximation holds for non-zero values of $V_\phi$, so one cannot use Eq. (\ref{eq:SReoms}) at $\phi = \phi_s$. Thus we will use the slow-roll equations for $\phi\neq\phi_s$. The cosmic inflation takes place as long as following slow-roll parameters are much smaller than one
\begin{equation}
\epsilon: = \frac{1}{2}\left(\frac{V_\phi}{V}\right)^2  \, , \qquad \eta := \frac{V_{\phi\phi}}{V} \ . \label{eq:SRparameters}
\end{equation}
The number of e-folds generated during the inflation is in the slow-roll approximation equal to 
\begin{equation}
N = \int_{t_i}^{t_f} \mathcal{H}d\tilde{t} \simeq \int^{\phi_i}_{\phi_f} \frac{V}{V_\phi}d\phi \, , \label{eq:Efolds}
\end{equation}
where indexes $i$ and $f$ refer to initial and final moments of inflation respectively. Namely, $t_i$ is the first moment when both slow-roll parameters are smaller than one and $t_f$ is the moment when any of slow-roll parameters become bigger than one. {In Fig. 6 of Ref. \cite{Artymowski:2015mva} we showed that during the saddle-point inflation {it is straightforward} to generate the number of e-folds significantly bigger than 60, even if the plateau is not present at all. }
\\*

The power spectra of the superhorizon primordial curvature perturbations and gravitational waves in the Einstein frame are following
\begin{equation}
\mathcal{P}_\mathcal{R} = \left(\frac{\mathcal{H}}{2\pi}\right)^2 \left(\frac{\mathcal{H}}{\phi'}\right)^2 \simeq \frac{V}{24\pi^2\epsilon} \, , \qquad \mathcal{P}_h = \left(\frac{\mathcal{H}}{2\pi}\right)^2. \label{eq:PP}
\end{equation}
The $\mathcal{P}_\mathcal{R}$ needs to be normalized at the horizon crossing of scales observed in the CMB. Usually one chooses the normalization moment to be at $N_\star \simeq 50-60$ (where the $\star$ denotes the value of the quantity at the horizon crossing), but in general the scale of normalisation strongly depends on the scale of inflation and reheating. In this {paper} we set $k_\star = 0.002 Mpc^{-1}$, where $k$ is the Fourier mode of perturbation, and $N_\star(\lambda_1 = \lambda_2 = 0) = 55$. The value of $N_\star$ decreases for big $\lambda_2$, since for lower scale of inflation the comoving Hubble radius will grow less during the post-inflationary era. This procedure sets $\mathcal{P}_\mathcal{R}^{1/2}\sim 5\times 10^{-5}$.
\\*

We have several parameters in the model, namely $\phi_\star$, $\phi_s$, $M$, $\lambda_1$ and $\lambda_2$, {however} since we require the existence of the saddle point, one finds $\lambda_2 = \lambda_2(\lambda_1,M)$ and $\phi_s = \phi_s (\lambda_1,M)$. From its definition $\phi_\star$ depends on $M$, $\lambda_1$ and $\lambda_2$. The normalization is a constraint which { we use to calculate required $M$, thus decreasing the amount of free parameters to just $\lambda_1$}. We will use $\lambda_1$ to parametrise the deviation from the Starobinsky inflation.
\\*

Two parameters of power spectra, which connect theory with the experiment are the tensor to scalar ratio $r$ and the scalar spectral index $n_s$ defined by
\begin{equation}
r = \frac{\mathcal{P}_h}{\mathcal{P}_\mathcal{R}} \simeq 16\epsilon \, , \qquad n_s = 1+ \frac{d\log\mathcal{P}_\mathcal{R}}{d\log k} \simeq 1 - 6\epsilon + 2\eta \, , \label{eq:rns}
\end{equation}
where $k$ is the Fourier mode. In the saddle point inflation one expects $\epsilon\ll \eta$, so $r_\star \ll 1$ and $n_{s\star} \simeq 1 + 2\eta_\star$. Since $n_{s\star} > 1$ is excluded by the experimental data \cite{Ade:2013uln} one needs $\eta<0$ at the moment of freeze-out. Thus, one requires $\phi_\star < \phi_s$. The total amount of e-folds produced for $\phi<\phi_s$ strongly depends on the length of the Starobinsky plateau and therefore on $\lambda_1$. If the plateau for $\phi<\phi_s$ is long enough to generate at least $60$ e-folds of inflation then the $\mathcal{P}_\mathcal{R}$ does not significantly deviate from the Starobinsky one. On the other hand for $\lambda_1\gg 1$ one may decrease the length of the plateau in order to generate e-folds only via the saddle point inflation. As we will show, this may be the way to decrease $M$ and therefore to obtain the low-scale inflation. 
\\*

\begin{figure}[h]
\centering
\includegraphics[height=6cm,bb=0 0 252 228]{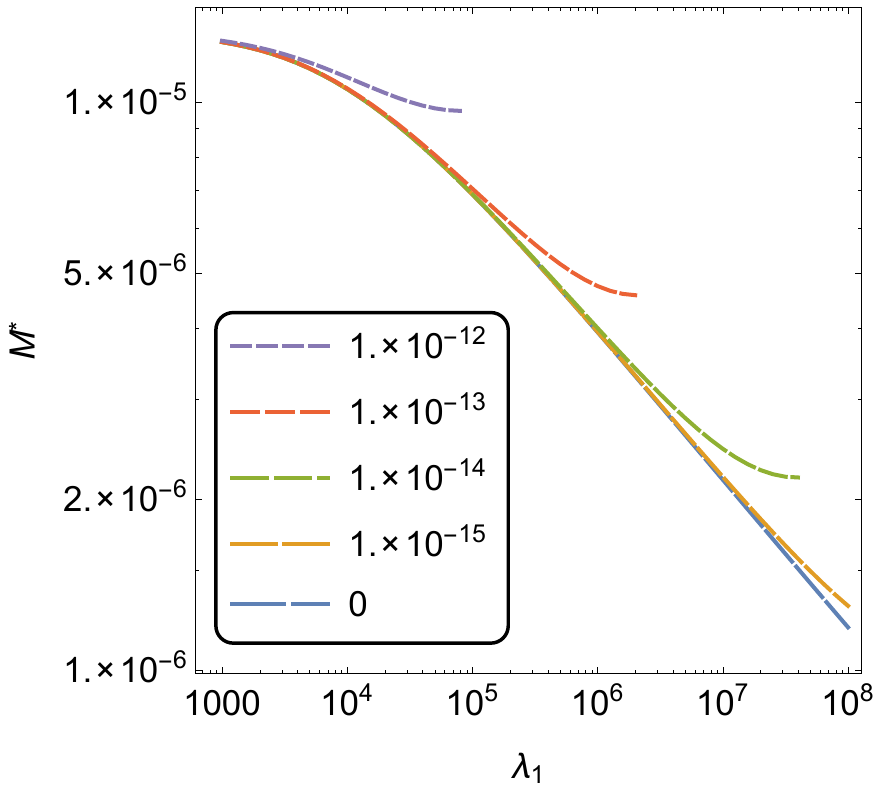}
\hspace{0.5cm}
\includegraphics[height=6cm,bb=0 0 252 252]{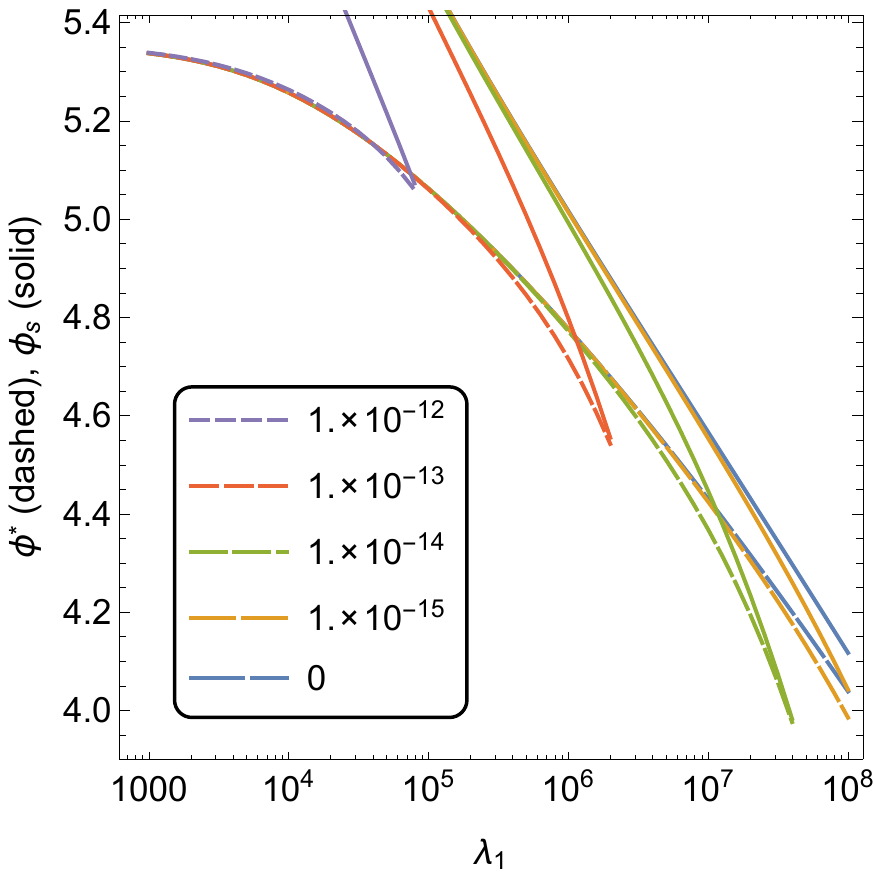}
\caption{\it Left Panel: $M(\lambda_1)$ {for several values of $V'(\phi_s)$. As expected, lower $V_{\phi}(\phi_s)$ gives lower scale of inflation}. Right panel: $\phi_\star$ and $\phi_s$ ({dashed and solid lines respectively}).} 
\label{fig:MPhiStar}
\end{figure}

\begin{figure}[h]
\centering
\includegraphics[height=5.7cm,bb=0 0 252 228]{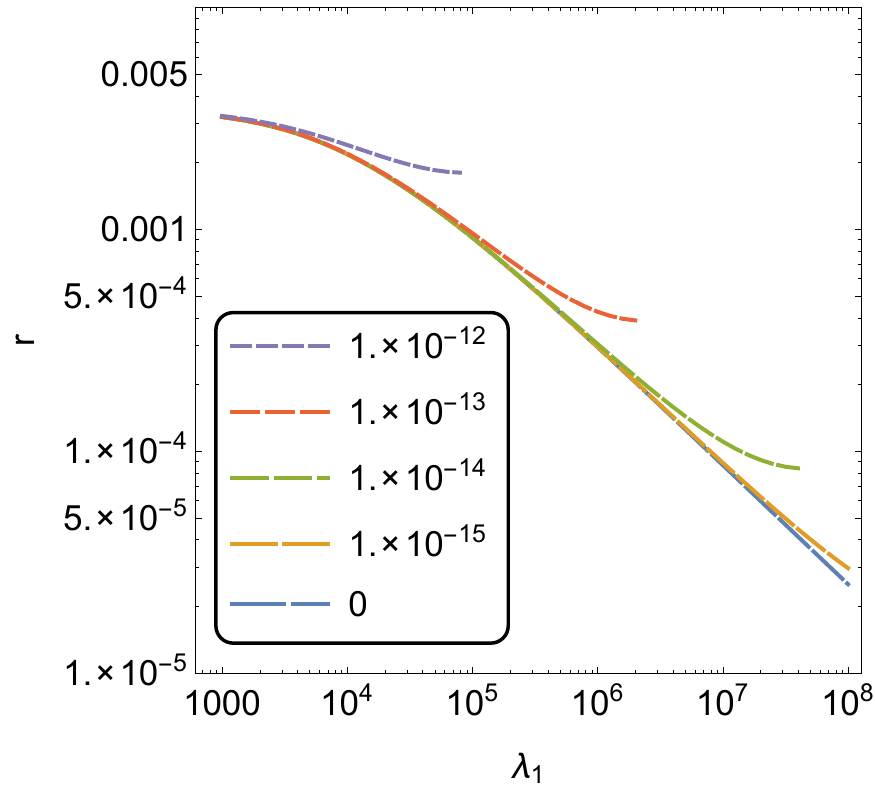}
\hspace{0.5cm}
\includegraphics[height=5.7cm,bb=0 0 303 295]{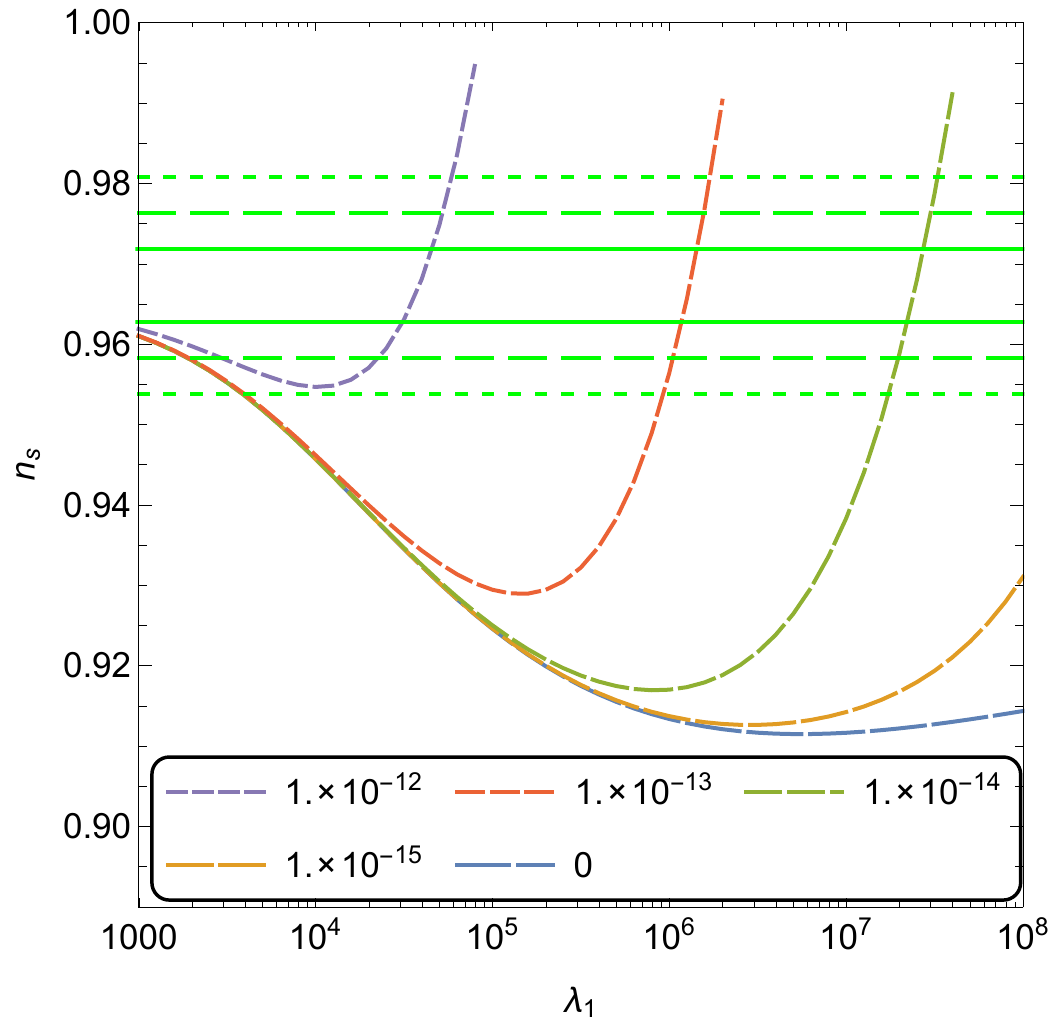}
\caption{\it Left panel: $r(\phi_\star)$ as a function of $\lambda_1$ for several values of $V_{\phi}(\phi_s)$. All obtained values of $r$ are consistent with the Planck data. For $\lambda_1\gtrsim 10^7$ one obtains $r\lesssim 0.002$, which gives $\Delta\phi<m_p$. Right panel: $n_s(\phi_\star)$ as a function of $\lambda_1$ for several values of $V_{\phi}(\phi_s)$. { One, two and three $\sigma$ regimes of the PLANCK 2015 best fit of $n_s$ lay between green lines (solid, dashed and dotted respectively).} {For $\lambda_1 \lesssim 10^3$ one obtains correct value of $n_s$ even for $V_\phi(\phi_s)=0$. In such a case last 60 e-folds of inflation are strongly influenced by the Starobinsky plateau.}}
\label{fig:rns}
\end{figure}

In Fig. \ref{fig:MPhiStar} we present $M$, $\phi_\star$ and $\phi_s$ as a function of $\lambda_1$. One can see that for $\lambda_1 > 10^{3}$ the $M$ decreases logarithmically with $\lambda_1$ to { reach $M\sim 10^{-6}$ at $\lambda_1 \sim 10^{8}$}. For $\lambda_1>10^6$ the normalisation of inhomogeneities happens very close to $\phi_s$. Nevertheless, for all considered values of $\lambda_1$ we obtained $\phi_\star < \phi_s$. In Fig. \ref{fig:rns} we present $r$ and $n_s$ for $\phi = \phi_\star$. {For $V_{\phi}(\phi_s) = 0$ (i.e. for a perfect saddle)} {and for $\phi_s\simeq\phi_\star$} one finds $n_s < 0.96$ and the model becomes inconsistent with the PLANCK data. {The perfect saddle may still fit the data for $\lambda_1\lesssim 10^3$}{, however then inflation simply occurs on the plateau far from the saddle and the whole modification is simply irrelevant. } 
 Thus let us consider the case of $V_{\phi\phi} = 0\, , V_{\phi} \neq 0$ to check if the inflection point inflation gives the correct values of $n_s$. The value of the field at the inflection will also be denoted as $\phi_s$. The results of numerical analysis for saddle and inflection point inflation for different $\lambda_1$ and $V_{\phi}(\phi_s)$ are presented in Fig. \ref{fig:MPhiStar} and \ref{fig:rns}. For the inflection point inflation one finds the maximal allowed $\lambda_1$ for which the number of e-folds generated for $\phi\leq\phi_s$ is equal to $N_\star$, which is the number of e-folds at the horizon crossing for the normalisation scale $k_\star$.

The key result is that inflection point inflation satisfies experimental constraints even for large values of $\lambda_s$. This means that it remains a valid solution, and one is not limited to very small modifications of the potential which would simply mean going back to the Starobinsky model. This unfortunately is not true for the pure saddle case in which inflation has to occur on the plateau and the potential can only be modified for very large field values.
\\*

\begin{figure}[h]
\centering
\includegraphics[height=5.7cm,bb=0 0 500 478]{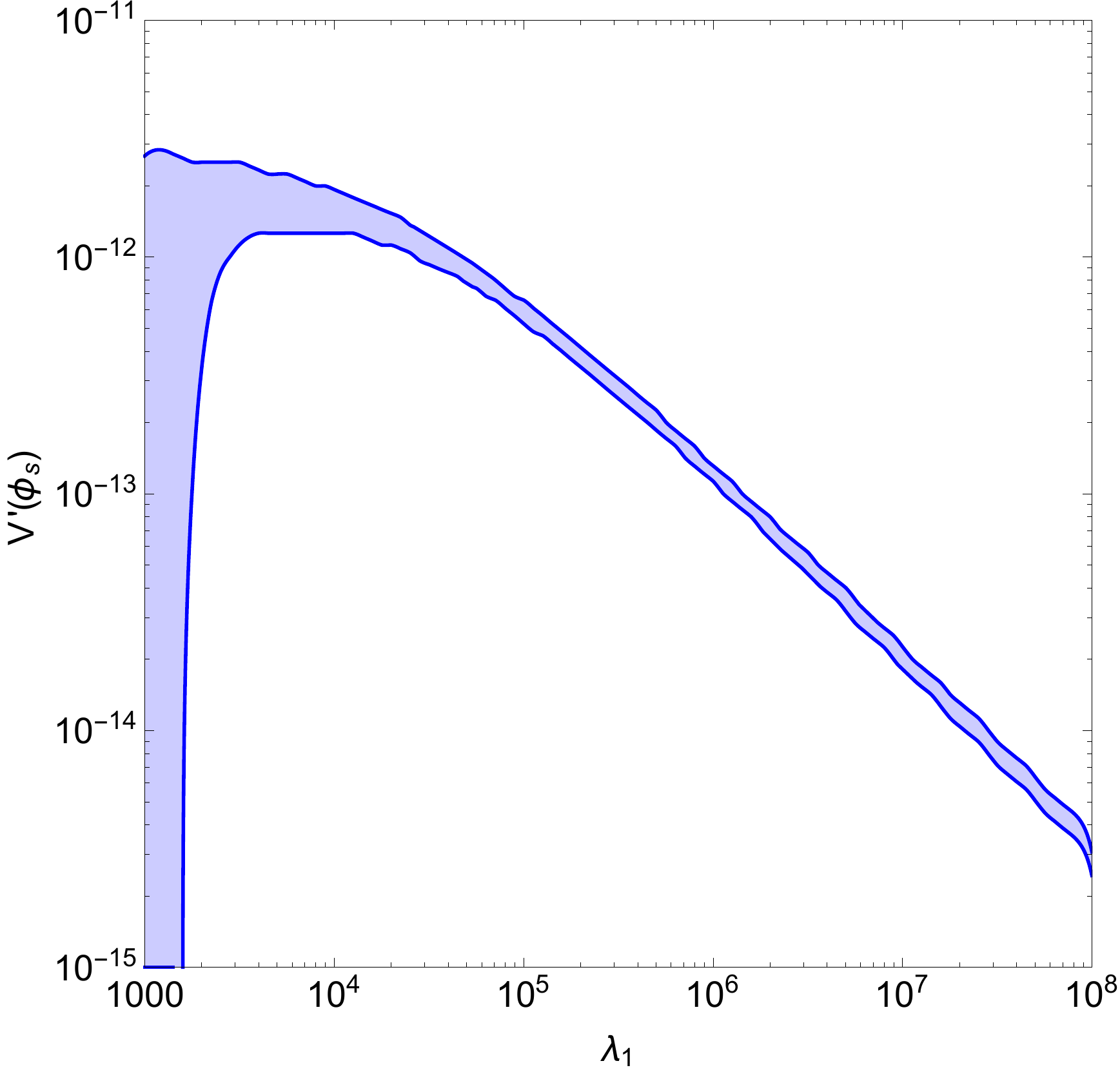}
\caption{\it The blue region represents the part of the $(\lambda_1,V_{\phi}(\phi_s))$ parameter space which fits $2\sigma$ constrains on $n_s$ from the PLANCK 2015 data. Smaller $V_{\phi}(\phi_s)$ corresponds to the smaller scale of inflation. One could extrapolate the blue region for $\lambda_1>10^8$, which would give $M^2 \lambda_1 \ll1$ for $\lambda_1<10^{20}$. For $\lambda_1 \lesssim 10^3$ any $V_\phi(\phi_s)$ can fit the data, including the perfect saddle point case, however this corresponds to inflation on the plateau far from inflection point}
\label{fig:Vplambda1}
\end{figure}


\section{Saddle point Higgs inflation}
\label{sec:higgs}

The saddle point inflation can be also obtained from higher order corrections to the so-called Higgs inflation. Let us define the action of a real scalar field (which in particular may be a Higgs field) with non-minimal coupling to gravity as
\begin{equation}
  S[\psi,g_{\mu\nu}] = \int d^4x\mathcal{L} 
  = \int d^{4}x\sqrt{-g}\left[U(\psi)R-\frac{1}{2}(\partial_{\mu}\psi)(\partial^{\mu}\psi)-W(\psi)\right] + S_m \, , \label{eq:Action}
\end{equation}
where $W(\psi)$ is the Jordan frame scalar potential, $R$ is the Ricci scalar, $U(\psi)$ is the function of non-minimal coupling to the gravity and $S_m$ is the action of matter fields, like dust, radiation, additional scalar fields etc. We assume that all fields in $S_m$ are minimally coupled to gravity. For $U \to 1/2$ one restores general relativity. In further parts of this section we will assume that
\begin{equation}
U(\psi)=\frac{1}{2}+\frac{1}{2}\xi\psi^2 \, ,\label{eq:Udef}
\end{equation}
where $\xi > 0$. Thus, the $\psi \to 0$ limit gives the GR vacuum. In the $W \propto \psi^4$ model the slow-roll parameters are proportional to $(\xi\psi^2)^{-1}$, so inflation happens for $\xi\psi^2\gg 1$. Let us assume that $W(\psi)$ is of the form
\begin{equation}
W(\psi) =  V_H + \frac{\lambda_6}{6\, M_p^2}\psi^6 + \frac{\lambda_8}{8\, M_p^4}\psi^8\, , \qquad \text{where} \qquad V_H = \frac{\lambda}{4} \psi^4 \, , \label{eq:Vmodified}
\end{equation}
where $\lambda_6,\lambda_8 = const$ and $W_H$ is a scalar potential which corresponds to the high energy approximation of the Mexican hat potential used e.g. to describe self-interaction of the Higgs field. Terms proportional to $\lambda_6$ and $\lambda_8$ are the higher order corrections to this potential \footnote{{The issue of saddle point Higgs inflation without higher order corrections has been partially analysed in Ref \cite{Bezrukov:2014bra}.}}. The $\lambda_i$ constants are dimensionless. As we will show one need to require $\lambda_8>0$ in order to obtain positive energy density and stability of the potential at very high energies. Nevertheless the sign of $\lambda_6$ remains undetermined. 
\\*

In order to obtain the canonical form of the action let us consider the transformation to the Einstein frame, namely
\begin{eqnarray}
 \tilde{g}_{\mu\nu} = 2U\, g_{\mu\nu}\, \qquad &\Rightarrow& \qquad  \tilde{a} = \sqrt{2U} a\ ,
  \qquad  d\tilde{t} = \sqrt{2U} d t\, , \label{eq:transformEin} \\
  \left(\frac{d\Psi}{d\psi}\right)^2 &=& \frac{1}{2}\frac{U+3U_\psi^2}{U^2}\, , \qquad V(\Psi) = \frac{W}{4U^2}(\psi = \psi(\Psi)) \, , \label{eq:phi}
\end{eqnarray}
where $\Psi$ is the Einstein frame field and $V$ is the Einstein frame potential. For $\xi\psi\gg1$ one finds $\psi \simeq \exp(\sqrt{2/3}\Psi)$, which is the result from the Brans-Dicke theory. The potential $V(\Psi)$ for several values of $\lambda_6$ and $\lambda_8$ parameters has been presented in Fig. \ref{fig:Saddle}. Let us note that for $\lambda_6 = \lambda_8 = 0$ one restores the potential from Ref. \cite{Bezrukov:2007ep}. Now the equations of motion take the form
\begin{eqnarray}
\Psi'' + 3\mathcal{H}\Psi' + V_\Psi = 0 \, , \qquad 3\mathcal{H}^2 = \frac{1}{2}\Psi'^2 + V \, ,
\end{eqnarray}
where $V_\Psi = \frac{dV}{d\Psi}$ and $\Psi' = \frac{d\Psi}{d\tilde{t}}$. The potential $V$ has a minimum at $\Psi = 0$, which corresponds to the GR limit of the theory. In the $\xi\gg1$, $\xi\psi^2 \gg 1$ approximation one finds following analytical relations, which {determine} the existence of the saddle point in the Einstein frame
\begin{equation}
\lambda_6 = -\lambda_s \sim 3\left(\frac{\lambda \lambda_8}{4\xi}\right)^{1/3} \, , \qquad \psi_s \simeq \sqrt{-\frac{\lambda _6}{3\lambda _8}}+ \frac{9\sqrt{3} \lambda  \lambda _8^{3/2}}{2 \left(-\lambda _6\right)^{5/2} \xi } \, .
\end{equation}
For $\lambda_6 >-\lambda_s$ the potential has a minimum at $\Psi = 0$ and flat plateaus, which ends with step, exponential slopes. The potential $V(\Psi)$ is always growing with $|\Psi|$ and one finds no stationary points. For $\lambda_6 = -\lambda_s$ the only stationary points besides the minimum at $\Psi = 0$ are two saddle points at $\Psi = \pm \Psi_s$. For $\lambda_6 < -\lambda_s$ the potential has additional minima and maxima at some $\Psi = \pm \Psi_{\min}$ and $\Psi = \pm \Psi_{\max}$ respectively.
\\*

\begin{figure}[h]
\centering
\includegraphics[height=5.7cm,bb=0 0 274 226]{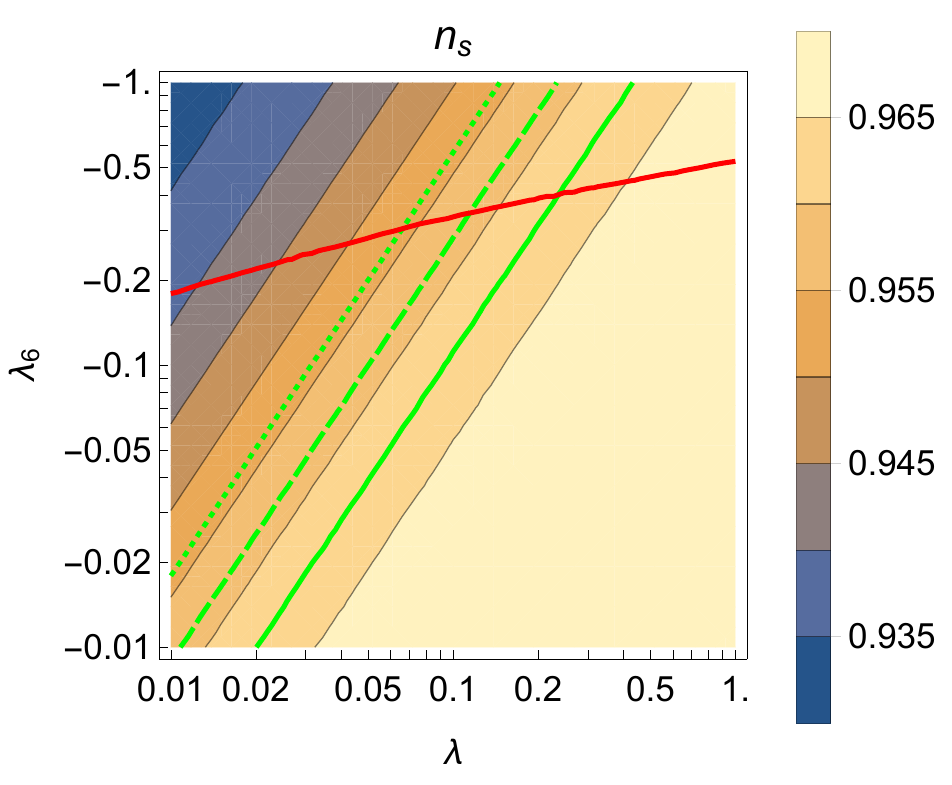}
\hspace{0.5cm}
\includegraphics[height=5.7cm,bb=0 0 280 226]{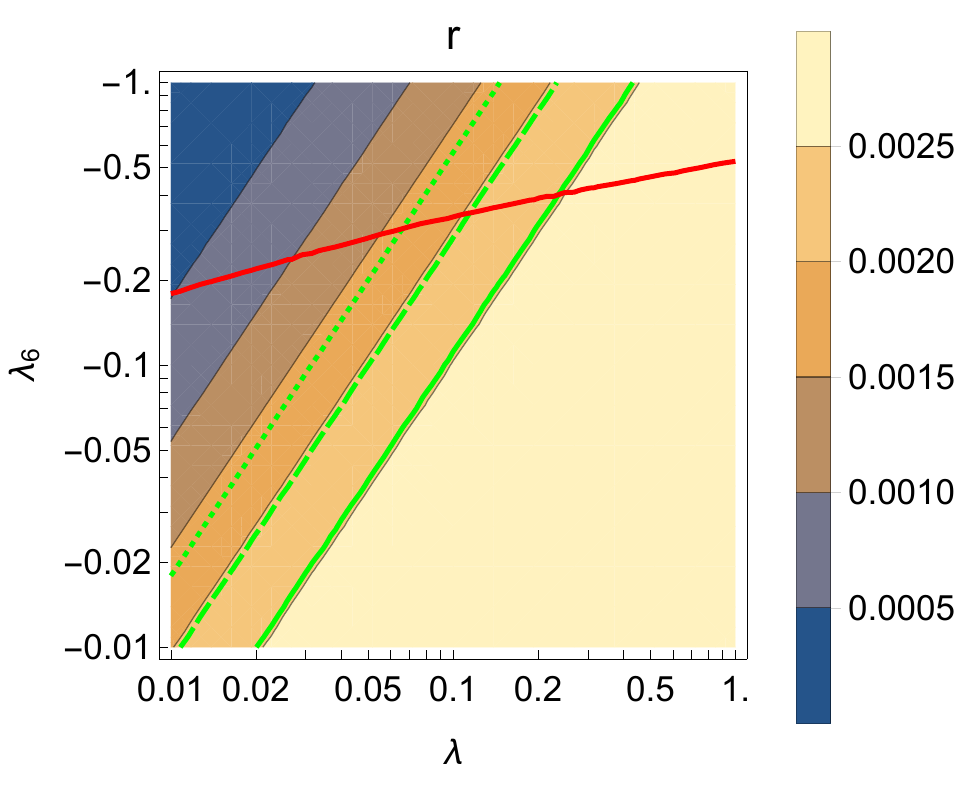}
\vspace{0.3cm}

\includegraphics[height=5.9cm,bb=0 0 266 226]{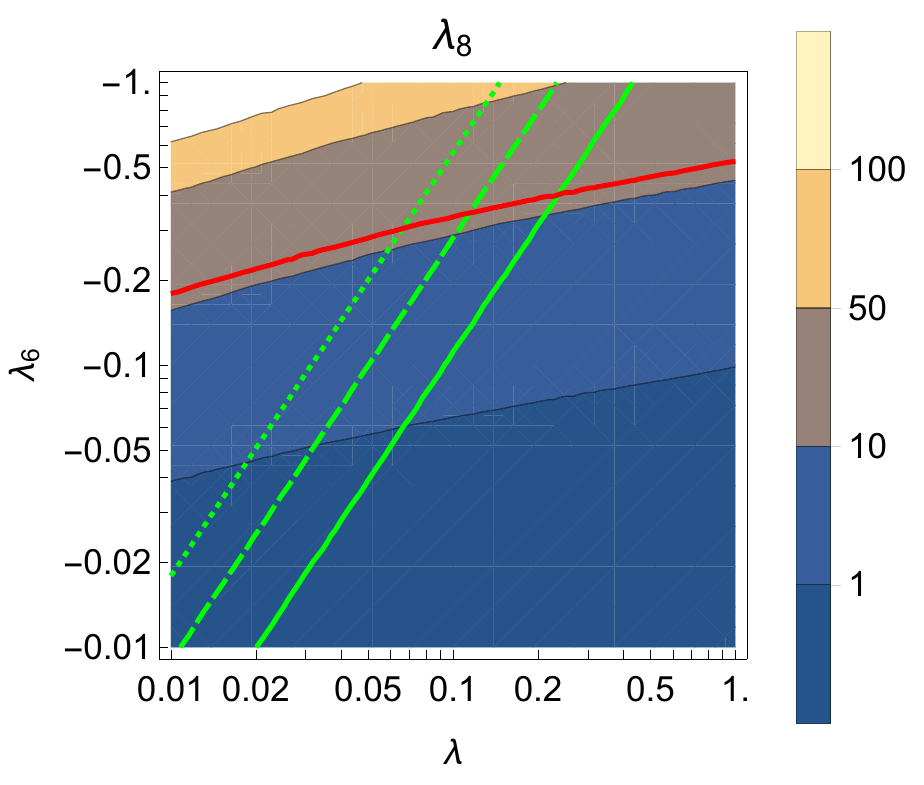}
\hspace{0.9cm}
\includegraphics[height=5.9cm,bb=0 0 263 226]{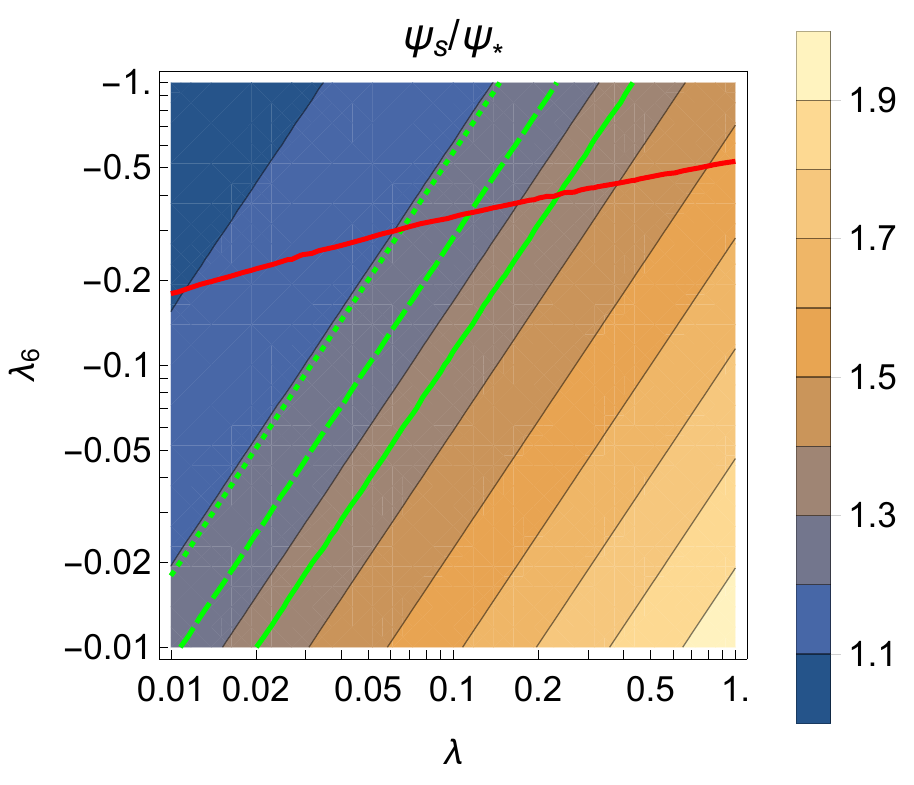}
\caption{\it {Left upper panel: spectral index $n_s=n_s(\lambda,\lambda_6)$. One, two and three $\sigma$ regimes of the PLANCK 2015 best fit of $n_s$ are below green lines (solid, dashed and dotted respectively){, which correspond to $\lambda_6 = -3.5\lambda^{3/2}$, $\lambda_6 = -8.8 \lambda^{3/2}$ and $\lambda_6 = 18\lambda^{3/2}$ respectively}. Right upper panel: tensor to scalar ratio $r$. All values of $r$ presented in the plot are consistent with PLANCK results. Left lower panel: $\lambda_8$ as a function of $\lambda$ and $\lambda_6$. In order to obtain perturbative theory one requires $\lambda_8< 10$, which gives $\lambda_6 < 0.1 \lambda^{1/4}$. The allowed region lies under the red line. Right lower panel: $\Psi_s/\Psi_\star$, which determinates how close to the saddle point one obtains freeze out of given scale $k_\star$.}}
\label{fig:Higgsrns}
\end{figure}

\begin{figure}[h]
\centering
\includegraphics[height=5.7cm,bb=0 0 279 226]{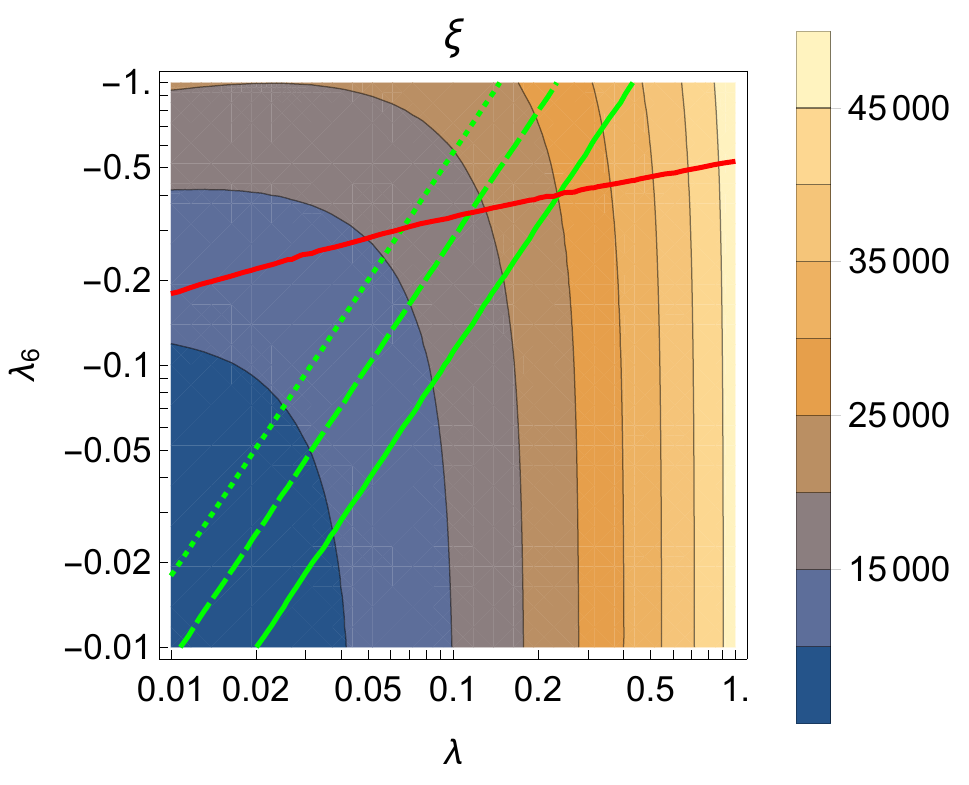}
\caption{\it The non-minimal coupling parameter $\xi$ with constrains on $\lambda_8$ and $n_s$. For $\lambda \gtrsim 0.3$ the $\lambda_6$ dependence is very weak and therefore one recovers $\lambda/\xi^2 = const$. Under the dashed green line (the PLANCK $2\sigma$ regime of $n_s$) the $\lambda_6$ dependence is weak for all $\lambda$ considered in the plot.}
\label{fig:Higgsxi}
\end{figure}

We start from 5 free parameters: $\lambda$, $\lambda_6$, $\lambda_8$, $\xi$ and $\Psi_s$. We have 3 constraints: $V_\Psi=0$, $V_{\Psi\Psi}=0$ and $\mathcal{P}_\mathcal{R}^{1/2}\simeq 5\times 10^{-5}$, which {we use to determine $\lambda_8$, $\xi$ and $\Psi_s$}. In Fig. \ref{fig:Higgsrns} we show the $\lambda$ and $\lambda_6$ dependence of $r$, $n_s$, $\lambda_8$ and $\Psi_s/\Psi_\star$. In Fig. \ref{fig:Higgsxi} we show weak $\lambda_6$ dependence of $\xi$ in the allowed part of the parameter space. {The $\lambda$ parameter is naturally limited by $4\pi$. In order to preserve perturbativity of the theory let us assume that $|\lambda_6|$ and $\lambda_8$ should not be bigger than $\mathcal{O}(1)$.} {It is crucial that even including all these constraints a significant part of the parameter space remains valid. Even though inflation occurs some distance away from the saddle ($\Psi_s / \Psi_* \gtrsim 1.26$) and so some influence of the plateau is inevitable, such an extension remains a viable extension of the standard Higgs inflation scenario.}

\section{Conclusions}\label{sec:concl}

In this paper we investigated the possibility of generating saddle point inflation from higher order corrections to theories of modified gravity, namely the Starobinsky model and Higgs inflation. It is crucial that even after including all constraints we discussed, a large portion of the parameter space in both models remains valid. Even though in modified Higgs inflation scenario and Starobinsky model with a pure saddle, inflation has to occur some distance away from the saddle to achieve correct values of $n_s$, and so some influence of the plateau is inevitable. Inflection or saddle point inflation remain a viable extension of the standard Starobinsky model and Higgs inflation scenarios.

The significant difference between the two is the number of free parameters which in the Higgs inflation case allows solutions with pure saddle to be consistent with all constraints. On the contrary, in modification of the Starobinsky model we had to resort to inflection point inflation rather than a pure saddle in order to achieve new valid results. {Note that higher order corrections do not change the position of Einstein frame vacuum and therefore they do not change the consistency between GR and low energy predictions of discussed theories.}
\\*

In Sec. \ref{sec:Infl} we introduced higher order corrections to the Jordan frame potential of the Starobinsky theory. We presented the analytical analysis of the existence of the saddle point of the Einstein frame potential.
\\*

In Sec. \ref{sec:infl2} we analyse features of primordial inhomogeneities generated during saddle point inflation. From the normalisation of perturbations and saddle (inflection) point conditions we obtained all parameters of the model (such as $M$, $n_s$, $r$ and $\phi_\star$) as functions of $\lambda_1$. We found the region in the $(\lambda_1,V_\phi(\phi_s))$ space which fits the $2\sigma$ regime of PLANCK. We showed how $r$ (and therefore the scale of inflation) can decrease for big $\lambda_1$ and we estimated the maximal allowed value of $\lambda_1$ to be of order of $10^{20}$. The inflation can happen almost exactly at the inflection point and no Starobinsky plateau is needed to obtain correct shape of the power spectrum. For $\lambda_1 \lesssim 10^3$ consistency with PLANCK can be obtained even for $V_\phi(\phi_s) = 0$. In such a case one obtains the Starobinsky plateau for $\phi<\phi_s$, which has significant influence on the generation of primordial inhomogeneities during the last 60 e-folds of inflation.
\\*

In Sec. \ref{sec:higgs} we investigated the  issue of higher order corrections to the $\lambda \psi^4$ potential. We found conditions for the existence of a saddle point as well as constrains in the $(\lambda,\lambda_6)$ plane, which come from perturbativity of the theory and consistency with PLANCK. The result is that the saddle point inflation is possible in such a model. However during the last 60 e-folds of inflation, interesting part of the potential is also influenced by the $\lambda \psi^4$ term, which generates Starobinsky-like plateau. This comes from the fact that in the allowed region of parameters $\Psi_s/\Psi_\star \gtrsim 1.26$, which means that inflation has to proceed on the plateau some distance from the saddle, in order to satisfy experimental constraints.

\section*{Acknowledgements}
This work was partially supported by the Foundation for Polish Science International PhD Projects Programme co-financed by the EU European Regional Development Fund and by National Science Centre under research grants DEC-2012/04/A/ST2/00099 and DEC-2014/13/N/ST2/02712.
ML was supported by the Polish National Science Centre under doctoral scholarship number 2015/16/T/ST2/00527. MA was supported by National Science Centre grant FUGA UMO-2014/12/S/ST2/00243. 


\end{document}